\documentstyle{article}
\begin{document}
\title{Non-Abelian Collective Excitations in Unlinearized Quark-Gluon Plasma
Media}
\author{{Zheng Xiaoping$^{1 \hskip 2mm 2}$\ \ Hou Defu$^{2}$\ \ Li
Jiarong$^{2}$}\\
{\small 1 Department of Physics,Zhongshan University,Guangzhou,P.R.China}\\
{\small 2 The Institute of Particle Physics, Huazhong Normal
University,Wuhan,P.R.China}}
\maketitle

\vskip-9.5cm
\hskip11cm{\large HZPP-9906}

\hskip11cm{\large July 25, 1999}

\vskip8.5cm

\begin{abstract}
\begin{minipage}{120mm}
We study the effect of unlinearized medium on the collective excitations in
quark-gluon plasma. We present two kinds of non-Abelian oscillation
solutions
which respectively correspond to weakly and strongly nonlinear coupling of
field components in color space. We also show that
the weakly nonlinear solution is similar to Abelian-like one but has the
frequency shift, which is of order $g^2T$, from eigenfrequency.
\vskip 0.5cm
 PACS number: 12.38.Mh
\end{minipage}
\end{abstract}
\vskip 0.9cm

  Collective excitations play an important role in the analyses of static
and dynamic properties of a many-particle system such as a plasma.
It is an extensively well-studied subject in electromagnetic
plasma\cite{r1,r2}.
The  quark-gluon plasma  oscillation modes have been studied for a long time
in the framework of finite-temperature(FT)QCD\cite{r3} and in kinetic-theory
approach\cite{r4}. First of all, most studies are limited to the
linearized equations for both Yang-Mills field equations and the kinetic
equations\cite{r4}, where the excitations reduce to Abelian-like plasma
waves.
Up to 1994, Blaizot and Iancu gave the non-Abelian oscillation
solutions at leading order in the gauge coupling $g$ approximation, where
  the medium still remains to be Abelian-like\cite{r5} although the mean
field
equations are unlinearized. The purpose of  this Letter is to present new
non-Abelian solutions that we obtained recently in true non-Abelian the
medium.

It is well known that the equations satisfied by the gauge mean fields
$A^\mu_a(x)$ in a medium
[Throughout this work, the greek indices refer to Minkovski space, while the
Latin subscripts indicate color indices for the generators of the gauge
group SU(N)] are
\begin{equation}
D^\mu F^a_{\mu\nu}=j^{{\rm ind}a}_\nu(x),
\end{equation}
where $D^\mu=\partial^\mu+igA^\mu(x), A^\mu\equiv A^\mu_a\tau^a$, and
$F^{\mu\nu}=\partial^\mu A^\nu-\partial^\nu A^\mu-igf_{abc}\tau^c
A^\mu_aA^\nu_b \tau_a$, the generator of SU(N).The energy density of
the collective excitation in the plasma can be written as
\begin{equation}
T^{00}(x)\equiv T^{00}_{\rm YM}(x)+t^{00}(x),
\end{equation}
with
$\partial_tt^{00}(x)=-{\bf E}_a\cdot{\bf j}^{{\rm ind}a}(x)$,
where $T^{00}(x)$ is the standard Yang-Mills contribution\cite{r6}.
The induced current $j^{\rm ind}_{\nu}$ describes the
response of the plasma to the color gauge fields $A^\mu_a$, which relates to
the fluctuations in the phase-space color densities of quarks and gluons.
It is expressed as
\begin{eqnarray}
j^{{\rm ind}a}_\nu&=&g\int {{\rm d^3p}\over (2\pi)^3}v_\mu
\{N_f[Q_+^{(n)a}({\bf
p},x)- Q_-^{(n)a}({\bf p},x)]\nonumber\\
&+&{\rm tr} [T_aG^{(n)}({\bf p},x)]\},
\end{eqnarray}
where
the quark( antiquark) and gluon distribution functions ($Q_{\pm}({\bf
p},x)$and $G({\bf p},x)$) are govern by the following
kinetic equations,which describe non-Abelian plasma medium\cite{r7}.
\begin{equation}
p^\mu D_\mu Q_{\pm}({\bf p},x)\pm {g\over
2}p^\mu\partial^\nu_p\{F_{\mu\nu}(x),Q_\pm ({\bf p},x)\}=0,
\end{equation}
\begin{equation}
p^\mu {\cal D}_\mu G({\bf p},x)+{g\over 2}p^\mu\partial^\nu_p\{{\cal
F}_{\mu\nu}(x),G({\bf
p},x)\}=0,
\end{equation}
where the calligraphic letters represent the corresponding operators
 in adjoint representation of SU(N), the generator denoted by $T_a$.
From these equations we can obtain the fluctuations around the average
phase-space color densities of plasma constituents. For convenience,
we choose the temporal gauge later on,
$A^0_a=0$. Therefore, for the solutions which are uniform in space we have
\begin{equation}
\partial_tQ^1_\pm =\mp g{\bf v}\cdot\partial_t{\bf A}{{\rm
d}Q^0_\pm\over{\rm d}p},
\end{equation}
\begin{equation}
\partial_tG^1 =-g{\bf v}\cdot\partial_t{\bf\cal A}{{\rm d}G^0\over{\rm d}p},
\end{equation}
\begin{eqnarray}
\partial_t Q^2_\pm &=&\mp ig^2[{\bf v\cdot A}, {\bf v\cdot A}]{{\rm
d}Q^0_\pm\over{\rm d}p}\nonumber\\
&\pm& {g^2\over 2}\{{\bf v}\cdot\partial_t{\bf A},{\bf v}\cdot{\bf A}\}{{\rm
d}^2Q^0_\pm\over{\rm d}p^2},
\end{eqnarray}
\begin{eqnarray}
\partial_t G^2&=&\mp ig^2[{\bf v\cdot{\cal A}}, {\bf v\cdot{\cal A}}]{{\rm
d}G^0\over{\rm d}p}\nonumber\\
&+& {g^2\over 2}\{{\bf v}\cdot\partial_t{\bf\cal A},{\bf v}\cdot{\bf\cal
A}\}{{\rm
d}^2G^0\over{\rm d}p^2},
\end{eqnarray}
\begin{eqnarray}
\partial_t Q^3_\pm &=&\pm g^3[{\bf v\cdot A}, \int {\rm d}t[{\bf v\cdot A},
{\bf v\cdot
A}]]{{\rm d}Q^0_\pm\over{\rm d}p}\nonumber\\
&+& {ig^3\over 4}[{\bf v\cdot A}, \{{\bf v\cdot A}, {\bf v\cdot A}\}]{{\rm
d}^2Q^0_\pm\over{\rm
d}p^2}\nonumber\\
&+&{ig^3\over 2}\{{\bf v}\cdot\partial_t{\bf A},\int{\rm d}t[{\bf
v}\cdot{\bf
A},{\bf v}\cdot{\bf A}]\}{{\rm d}^2Q^0_\pm\over{\rm d}p^2}\nonumber\\
&\mp& {g^3\over 8}\{{\bf v}\cdot\partial_t{\bf A}, \{{\bf v\cdot A}, {\bf
v\cdot A}\}]{{\rm
d}^3Q^0_\pm\over{\rm d}p^3},
\end{eqnarray}
\begin{eqnarray}
\partial_t G^3 &=&+ g^3[{\bf v\cdot{\cal A}}, \int {\rm d}t[{\bf v\cdot
{\cal A}}, {\bf v\cdot{\cal A}}]]{{\rm d}G^0\over{\rm d}p}\nonumber\\
&+& {ig^3\over 4}[{\bf v\cdot{\cal A}}, \{{\bf v\cdot{\cal A}}, {\bf
v\cdot{\cal A}}\}]{{\rm d}^2G^0\over{\rm d}p^2}\nonumber\\
&+&{ig^3\over 2}\{{\bf v}\cdot\partial_t{\bf\cal A},\int{\rm d}t[{\bf
v}\cdot{\bf
\cal A},{\bf v}\cdot{\bf\cal A}]\}{{\rm d}^2G^0\over{\rm d}p^2}\nonumber\\
&-& {g^3\over 8}\{{\bf v}\cdot\partial_t{\bf\cal A}, \{{\bf v\cdot\cal A},
{\bf
v\cdot\cal A}\}]{{\rm d}^3G^0\over{\rm d}p^3},
\end{eqnarray}

All prevous studies in  this field were limited to  linear induced current
in the linear response approximation\cite{r5,r6,r8,r9}. From Eq(3),(6) and
(7),
 we easily write
the  current as\cite{r5,r6}
\begin{equation}
{\bf j}^l_a=-\omega_p^2 {\bf A}^a(t),
\end{equation}
where $\omega_p^2\equiv {1\over 3}{g^2\over 2\pi^2}\int {\rm
d}pp^2{\partial\over\partial p}[N_f(Q^0_++Q_-^0)+2NG^0]$ is the plasma
frequency when $Q^0_\pm$ and $G^0$ take the Fermi-Dirac and Bose-Einstein
distributions respectively.
As we
have  argued above, the result in the linear approximation is Abelian-like.
We need to consider non-linear response of the plasma in order to get
non-Abelian effect of the medium. Therefore, we also derive the non-linear
induced currents from
the non-linear density fluctuations (8),(9),(10) and (11). However, we are
able to
neglect the second-order current in contrast to the third-order
current\cite{r10}
(In fact, we can exactly prove that the
even-order currents are always zero). Thus, the third-order current is the
 non-linear leading order contribution(We only consider the lowest order
 here), which is calculated according to the formulae below
\begin{eqnarray}
{\bf j}^n_a&=&{\rm tr}(2\tau^a{\bf J}_Q)+{\rm tr}(T^a{\bf J}_G),\nonumber\\
{\bf J}_{Q}&=&{\bf J}_{Q_+}-{\bf J}_{Q_-}\nonumber\\
&=&\int {{\rm d}^3p\over (2\pi)^3}{\bf v}N_f(Q^3_+-Q^3_-),\nonumber\\
{\bf J}_G&=&\int {{\rm d}^3p\over (2\pi)^3}{\bf v}G^3.
\end{eqnarray}

Both  linear and non-linear density fluctuations contributing to field
equations(1) and energy density(2),  we have
\begin{equation}
 {{\rm d}^2A_i\over {\rm d}t^2}+\omega_p^2A_i+g^2[[A_i,A_j],A_j]+j^n_i=0,
\end{equation}
and
\begin{eqnarray}
T^{00}(t)&=&{1\over 2}\left ({{\rm d}{\bf A}_a\over{\rm d}t}\cdot {{\rm
d}{\bf
A}_a\over{\rm d}t}+\omega^2_p{\bf A}_a\cdot{\bf A}_a+\int{\rm
d}t{{\rm d}{\bf A}_a\over{\rm d}t}\cdot{\bf j}^n_a\right )\nonumber\\
&+&{g^2\over 4}f^{abc}f^{ade}({\bf A}_b\cdot{\bf A}_d)({\bf A}_c\cdot{\bf
A}_e).
\end{eqnarray}
Note that $T^{00}$ is a conserved quantity and manifestly positive.
We restrict ourselves to such configurations for the case of SU(2), and
assume, for simplicity, that $A^i_a(t)={\cal C}^i_ah^i(t)$(no summation over
$i$), taking constant${\cal C}=\delta^i_a$ as did in previous works\cite{r5,r6}
.
This ansatz is what proposed by Baseyan, Matinyan and Savvidy\cite{r11}. We
calculate and
immediately find only the last terms of the right-hand side of
equations (10)and (11) contribute to the current ${\bf j}^n_a$ under the
ansatz.
We obtain
\begin{equation}
{\bf j}^n_a=-{g^2\over 2}\omega^2_{p'}(3h^3_i+h_i\sum\limits_{j\not
=i}h_j^2),
\end{equation}
where $\omega^2_{p'}={g^2\over 180}\int{p^2{\rm d}p\over
2\pi^2}{\partial^3\over\partial p^3}[N_f(Q^0_++Q^0_-)+8NG^0]$ represents
non-linear collective effect of plasma.
The field equations(14) then become
\begin{eqnarray}
{{\rm d}^2h_i\over{\rm d}t^2}&+&\omega_p^2h_i+g^2h_i\sum\limits_{j\not=
i}h^2_j\nonumber\\
&+&{g^2\over 2}\omega^2_{p'}(3h^3_i+h_i\sum\limits_{j\not =i}h^2_j)=0,
\end{eqnarray}
These equations contain not only the linear but also non-linear
  thermal mass terms. They are similar to the past equations  coupled
nonlinearly with each other. However,it is more important that they  differ,
as we shall see, by
the presence of self-coupled thermal term ${g^2\over 2}\omega^2_{p'}3h^3_i$
 from the effect of the non-linear response of plasma. The  energy density
associated with $h_i$ is
\begin{eqnarray}
T^{00}&=&{1\over 2}\sum\limits_i\left [\left ({{\rm d}h_i\over{\rm
d}t}\right
)^2+{g^2\over 2}h^2_i\sum\limits_{j\not =i}h^2_j\right.\nonumber\\
&+&\left.\omega_p^2h_i^2+{g^2\over 2}\omega^2_{p'}({3\over
2}h^4_i+h^2_i\sum\limits_{j\not
=i}h^2_j)\right ],
\end{eqnarray}
where the effective Hamiltonian is composed of two terms of field
contributions
and two terms of thermal effects.

We see that the simplest motion is one dimensional. Let us consider now
the solutions to equations (17) for the motion. Then the equations(17)
reduce to
for $h_1=h, h_2=h_3=0$
\begin{equation}
{{\rm d}^2h\over{\rm d}t^2}+\omega_p^2h
+{g^2\over 2}\omega^2_{p'}3h^3=0,
\end{equation}

We discuss now it for two cases: nearly equilibrium and far from
equilibrium.

(i) First of all, we rewrite the non-linear self-coupled term in(19) as
${g^2\over 2}\omega^2_{p'}3\langle h^2\rangle h$, where we replace $h^2$ by
its time
average $\langle h^2\rangle$ which is so called fluctuation correlation
intensity.
We know correlation is small at the nearby equilibrium state of system.
Therefore, the non-linear term  is smaller than the linear term in(19)
because
the self-coupled term is two higher orders in the coupling constant $g$
than the linear term. We replace then equation(19) by the following
Iterating
equations

\begin{equation}
{{\rm d}^2h^{(0)}\over{\rm d}t^2}+\omega_p^2h^{(0)}=0,
\end{equation}

\begin{equation}
{{\rm d}^2h\over{\rm d}t^2}+[\omega_p^2
+{g^2\over 2}\omega^2_{p'}3(h^{(0)})^2]h=0,
\end{equation}
We have oscillation solution to the equations for $\phi=0$ at initial time
as follows
\begin{equation}
h^{(0)}(t)=\sqrt 2\theta\cos{\omega_pt},
\end{equation}
\begin{eqnarray}
h(t)=h^w_\theta\cos{[\omega_\theta t(1+{3\over
4}\omega^2_{p'}\theta^2)+{3\over
8}\omega^2_{p'}\theta^2\sin 2\omega_pt]},
\end{eqnarray}
where $\theta^2={g^2T^{00}\over\omega^4_p}$ is the dimensionless
Hamiltonian,
$h^w_\theta=\sqrt{2}\theta(1+{3\over2}\omega^2_{p'}\cos^4)^{1\over
2}\omega_pt$.
$h(t)$is weakly nonlinear and there exists an frequency shift from the
eigenvalue
because of nonlinear effect of  plasma
\begin{eqnarray}
\Delta\omega&=&\omega_\theta-\omega_p\nonumber\\
&=&{3\over 4}\omega_{p'}^2
\theta^2(1+{1\over 2\omega_pt}\sin{2\omega_pt})\omega_p,
\end{eqnarray}
We have
$\Delta\omega={3\over 4}\omega_{p'}^2
\theta^2\omega_p$
to be of order $g^2T$ coincided with previous works\cite{rr} when $\langle
(h^{(0)})^2\rangle$ replaces $(h^{(0)})^2$.

(ii)For far from equilibrium state, the  self-coupled nonlinear term in (19)
shall not be small when $\langle h^2\rangle$ is large enough.
In fact, it is possible to let $\langle h^2\rangle\gg 1,
\omega^2_{p'}\langle h^2\rangle\sim 1$, such as
turbulent plasma\cite{r12}. Of course, one expects there exist turbulent
QGP,
which has been roughly discussed in other works\cite{r10},
in heavy ion collisions\cite{r9}.  Equation(19) is strong nonlinear for
the case and we know a Jacobi elliptic cosine solution to it,
\begin{equation}
h(t)=h^s_\theta{\rm cn}[\Gamma^{-{1\over 2}}(2\theta^2+\Gamma^2)^{1\over
4}(\omega_pt-\phi);k],
\end{equation}
with the $h^s_\theta={\omega_p\over
g}[\Gamma(\sqrt{2\theta^2+\Gamma^2}-\Gamma)]^{1\over 2}$, the modulus
$k={1\over\sqrt 2}(1-{\Gamma\over\sqrt{2\theta^2+\Gamma^2}})^{1\over 2}$
and a period
${\cal T}={4\over \omega_p}{\Gamma^{1\over
2}\over{(2\theta^2+\Gamma^{2})}^{1\over 4}}K(k)$,
where $\Gamma^2={2\over 3\omega^2_{p'}}$ and $K(k)$ is the complete elliptic
integral modulus $k$.
What we have obtained here is an one dimensional solution similar to the two
dimensional case in ref\cite{r5}. The factor $\Gamma$ characterizes the
nonlinearity
and the non-Abelian property of the solution. The solution reduces to the
Abelian-like harmonic oscillation\cite{r5,r6} and
 ${\cal T}\rightarrow {\cal T}_0=2\pi/\omega_p$ when $\theta\ll 1$,
  $k\rightarrow 0$ and ${\rm cn}\rightarrow \cos$.

 In summary, we have studied here the non-Abelian oscillations considering
 the nonlinear response of the plasma. We have shown that self-coupled
 thermal term play a vital role for non-Abelian excitations. We have
obtained
   nonlinear solutions for SU(2) in one dimensional color space.
  We have also found a weak nonlinear solution similar to the Abelian-like
  harmonic oscillation but having a frequency shift from eigenfrequency
$\omega_p$
  for small correlation intensity and a strong nonlinear solution ,which is
  the typical non-Abelian excitation\cite{r5,r6}, for the inverse case. As
  Blaizot and Iancu\cite{r5} have pointed out, such as solutions for SU(2)
can be
  embedded in any larger SU(N) theory in the standard way\cite{r13}.

\end{document}